# The Heart of Protein-Protein Interaction Networks

Luciano da Fontoura Costa[1]

The recent developments in complex networks *(1–2)* have paved the way to a series of important biological insights.  An especially interesting finding *(3)* is the fact that a significant part of the mapped essential proteins of *S. cerevisae* corresponds to the so-called *hubs* of the scale-free complex networks obtained from ever growing protein-protein interaction data.  Characterized by high connectivity, hubs are particularly important nodes for several complex networks.  However, there are other types of network nodes, such as those corresponding to the network border (e.g. the nodes with only one connection), as well as the *innermost* nodes, which also deserve special attention.  This work reports on how the application of the concept of distance transform *(4)* to complex networks showed that a great deal of the network innermost nodes correspond to essential proteins, with interesting biological implications.

While the network boundary (or border) corresponds to the set of nodes with unit degree, it is possible to identify the network *innermost* nodes as those which are most distant from such a boundary.  Such nodes, characterized by at least two connections, are particularly important because, in several network growing models, they are likely to correspond to the oldest nodes incorporated during the network evolution, while the boundary nodes tend to be more recent. The identification of the network innermost nodes can be obtained through the distance transform of the network, performed by using an extension of the algorithm described in *(5)*. As hubs are not necessarily the oldest nodes *(1)*, the identification of the network border and innermost nodes can provide valuable complementary information about the evolution of the network and the role of its respective nodes.

We applied the above discussed concepts and algorithms to the *S. cerevisae* protein-protein interaction network described in *(3)*.   The *M* nodes with the highest distance values were identified in each case, and the number *R* of such nodes corresponding to essential proteins *(3)* was calculated and shown in Figure (a).  Considering that the expected proportion of essential proteins of a randomly assembled network would be 0.24, i.e. the number of essential proteins divided by the total number of proteins, the results in Figure 1(a) indicate that a significant percentage of the nodes most distant from the network border tend to correspond to essential proteins, while those closer to the border are much less likely to do so.  Figure 1(b) shows the subnetwork defined by all the 255 essential proteins, with the nodes which are also among the 300 innermost network nodes shown with larger diameter.  Interestingly, the dominant cluster,

[1] Institute of Physics at São Carlos, University of São Paulo, São Carlos, SP, Brazil, luciano@if.sc.usp.br. The author is grateful to H. Jeong for supplying the essential protein data, and to the *Human Frontier Science Program* for financial support.

emphasized by thicker edges, contains a particularly high ratio of innermost nodes.

The fact that several of the proteins which lie at the heart of the respective interaction network tend to be essential can be interpreted as an indication that they correspond to the oldest components of the biological cycles required to sustain life *(6)*. This hypothesis is further substantiated as most clusters of essential proteins incorporate at least one innermost node (see Fig. 1b). For such reasons, it is suggested that the proteins which are both essential and innermost would be especially important along the phylogenetic scale.

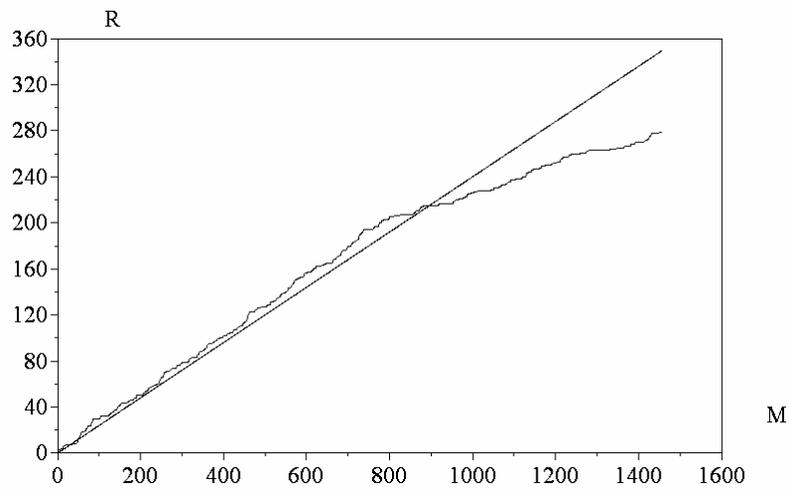

(a)

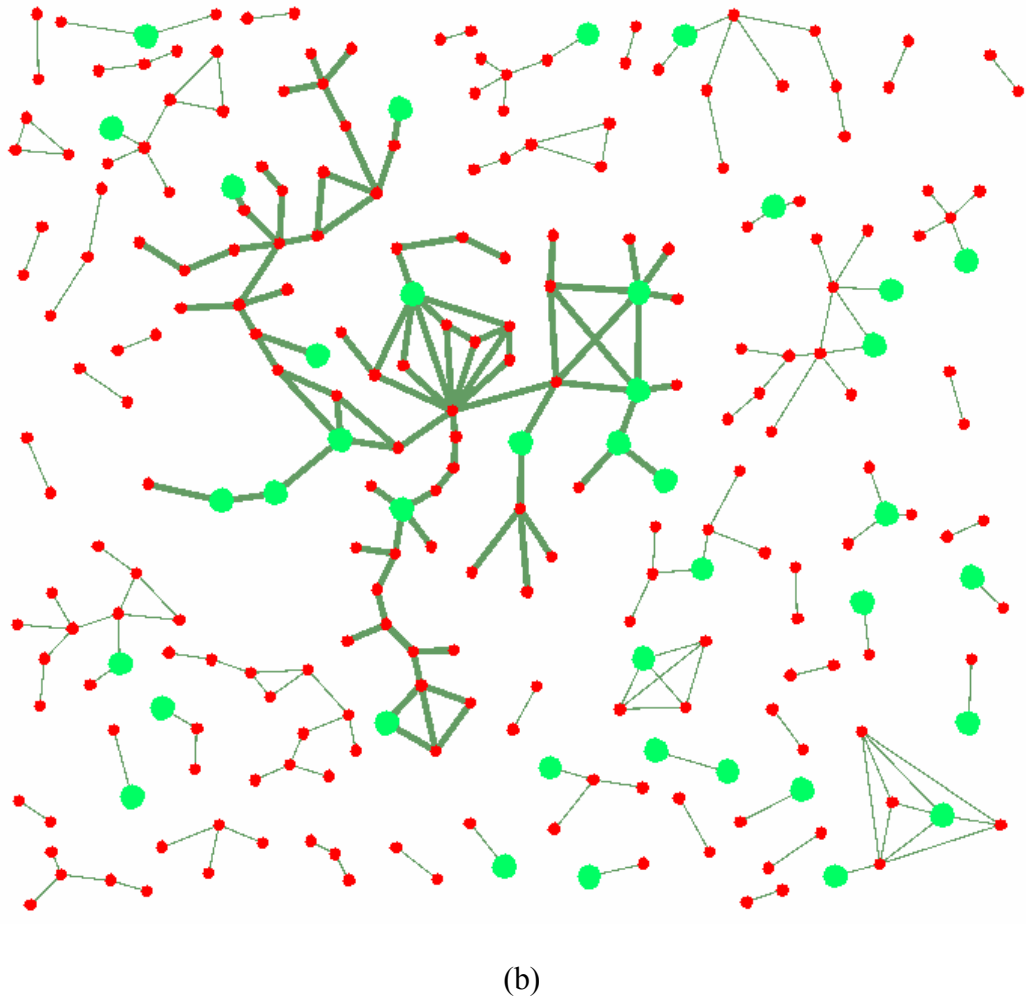

(b)

***Figure 1:*** The total $R$ of essential proteins for the $M$ innermost nodes, in decreasing order of distance from the network border (a). The straight line indicates the number of essential proteins which would be expected in case the essential nodes were uniformly distributed among the original network. The subnetwork defined by the 255 essential proteins is shown in (b), where the innermost nodes of the original protein-protein interaction network are shown in with larger diameters (green).

**ABSTRACT**

Recent developments in complex networks have paved the way to a series of important biological insights, such as the fact that many of the essential proteins of S. cerevisae corresponds to the so-called hubs of the respective protein-protein interaction networks. Despite the special importance of hubs, other types of nodes such as those corresponding to the network border, as well as the innermost nodes, also deserve special attention. This work reports on how the application of the concept of distance transform to networks showed that a great deal of the innermost nodes correspond to essential proteins, with interesting biological implications.